# QUENCH IN HIGH TEMPERATURE SUPERCONDUCTOR MAGNETS*

J. Schwartz, Department of Materials Science and Engineering, North Carolina State University, Raleigh, NC 27695, USA

*Abstract*

High field superconducting magnets using high temperature superconductors are being developed for high energy physics, nuclear magnetic resonance and energy storage applications. Although the conductor technology has progressed to the point where such large magnets can be readily envisioned, quench protection remains a key challenge. It is well-established that quench propagation in HTS magnets is very slow and this brings new challenges that must be addressed. In this paper, these challenges are discussed and potential solutions, driven by new technologies such as optical fiber based sensors and thermally conducting electrical insulators, are reviewed.

## INTRODUCTION

High temperature superconductor (HTS) wire technology has matured sufficiently such that there is an increasing effort to use them to generate high magnetic fields for applications including high energy physics, energy storage and nuclear magnetic resonance [1]. Of the superconductors discovered since 1987, two conductor technologies have emerged with the potential to generate magnetic fields well above 25 T, the limit of low temperature superconductor (LTS) technologies. One of these options, Ag/Ag-alloy clad $Bi_2Sr_2CaCu_2O_x$ (Bi2212), is the only HTS round wire (RW) with high critical current density ($J_c$) in magnetic fields above 20 T [2-6]. The manufacturing of Bi2212 RW has advanced sufficiently for the construction of insert coils that generate magnetic fields above 25 T [6-9]. The primary limitation of Bi2212 RW is a lack of mechanical strength. The other options, $(RE)Ba_2Cu_3O_{7-z}$ (REBCO) coated conductors, are comprised of a thin layer of REBCO superconductor deposited on a complex Ni-alloy substrate with one or more oxide buffer layers and subsequently encased in Cu. REBCO coated conductors have very high $J_c$ in the REBCO layer and very high mechanical strength in axial tension due to the Ni-alloy substrate, but are only available in a wide tape geometry. Although cabling options have been proposed, none clearly satisfies the requirements for cost-effective high field magnets [10, 11].

To safely operate a large, high field superconducting magnet, regardless of conductor type, a thorough understanding of the quench behavior of the conductor and magnet is required. Although the quench behavior of LTS-based magnets is well understood, HTS magnets show distinctly different quantitative behavior [12, 13]. For example, the minimum quench energy (MQE) in HTS magnets is quite high but the corresponding normal zone propagation velocity (NZPV) is a few orders of magnitude slower than in LTS magnets [13-23]. Thus, while the basic physics of quench behavior is unchanged, from a practical perspective the quantitative differences in behavior dominate.

The aim of any quench protection system is to prevent permanent conductor degradation in the event of a fault condition that induces a quench. As illustrated in Fig. 1, quench protection involves three key steps, all of which must be accomplished within a time-budget determined by the rate of growth of the disturbance and the resilience of the conductor. The three key steps are: (1) detection of a disturbance (historically accomplished with voltage measurements), (2) assessment of the disturbance to determine if it is going to induce a quench and to prevent false-positives, and (3) a protective action to prevent degradation if the magnet is quenching. For large magnets with a large stored energy, this is typically accomplished through heaters embedded in the magnet and a dump circuit into which the stored energy is dissipated. Thus, to develop an effective quench protection system, one must understand the stability and quench dynamics of the magnet in order to design a detection system, and one must also know the safe operational limits of the conductor. These two factors determine the time budget for protection. For LTS magnets, the safe operational limits are quantified in terms of a maximum hot-spot temperature; it is unclear if this is a sufficient criterion for HTS magnets.

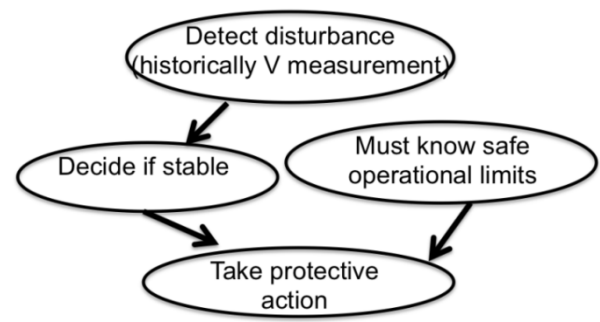

Figure 1: Illustration of quench protection workflow.

In considering the quench behavior of high field HTS magnets, it is important to recognize that such magnets are likely to be LTS/HTS hybrid systems, with LTS outsert coils that generate around 18 T and HTS insert coils that generate the highest magnetic fields [1]. Thus, other than during field ramp-up (in single power supply magnet systems), the HTS conductor will only be exposed to high magnetic fields and the behavior at low field is less important. Although most studies on HTS quench behavior have focused primarily on the behavior at self-field or relatively low magnetic field [24, 25], one recent study focused on the effects of high magnetic fields on quench behavior in Bi2212 RW coils [26]. The primary question was whether high magnetic field would increase the NZPV due to reduced critical temperature ($T_c$) and thus reduced current sharing temperature ($T_{cs}$) and temperature margin, or instead whether the NZPV would decrease due to reduced $J_c$. Fig. 2 plots the magnetic field dependence of the NZPV, $I_c$, MQE, $T_c$ and $T_{cs}$ for two Bi2212 coils and strands. While the MQE decreases with increased magnetic field for magnetic fields up to 20 T (the maximum reported), the NZPV is magnetic-field independent for fields greater than about 10 T. Thus, increased magnetic field results in reduced stability margin but not in enhanced propagation.

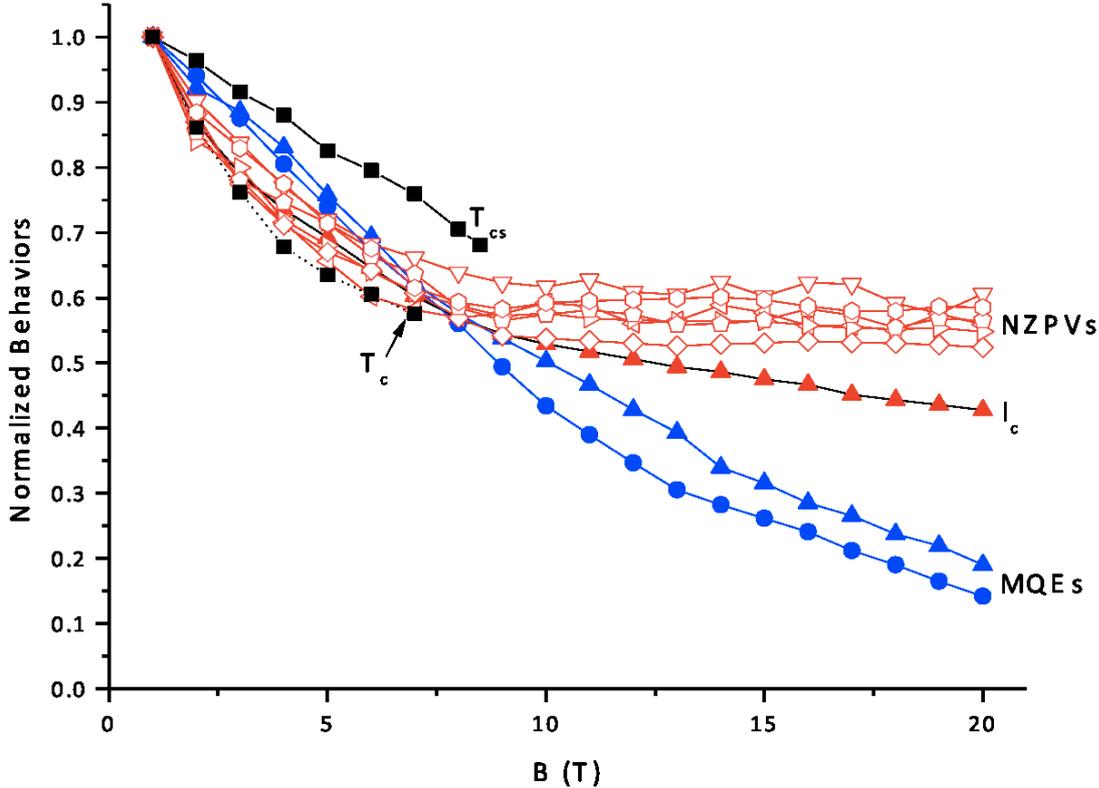

Figure 2: Normalized magnetic field-dependent quench behaviors [26].

## DEGRADATION IN HTS CONDUCTORS

The key to quench protection is preventing degradation by limiting localized temperature growth relative to the ability to detect and protect within the time available before degradation occurs. To determine the time budget, one must understand both the quench dynamics and the operational limits of the conductor. The safe operational limits for Bi2212 and REBCO conductors, however, are likely to not only be quite different from the limits on LTS conductors, but also from each other. Bi2212 RWs have defect-dominated microstructures with a significant quantity of interfilamentary bridges, porosity, non-superconducting phases, and $Bi_2Sr_2CuO_z$ intergrowths within the Bi2212 grains. Thus, electrical transport in Bi2212 RWs is not well understood, and identifying microscopic causes of reduced electrical performance is particularly challenging [3, 4]. Unlike Bi2212, REBCO coated conductors have highly engineered microstructures with very few unintended defects. Often non-superconducting, nanoscale defects are intentionally grown within the REBCO phase for enhanced flux pinning, but nonetheless the REBCO remains highly dense and relatively homogeneous [27-30].

### Bi2212 RWs

While the mechanisms of Bi2212 degradation are not understood, two studies have investigated the safe operational limits of Bi2212 mulitflamentary tapes and RWs [24, 31]. These studies have induced quenches in short strands and small test coils and allowed the quench to proceed systematically until the conductor degrades. The temperature-time data was interpreted in terms of the maximum temperature ($T_{max}$), the maximum temperature gradient ($dT/dx|_{max}$) and the maximum rate of temperature increase ($dT/dt|_{max}$); information about the conductors is seen in Table 1 and results are seen in Table 2. The results indicate that thermal shock ($dT/dt|_{max}$) is not the underlying driver for degradation. Furthermore, it is difficult to differentiate between the effects of $T_{max}$ and $dT/dt|_{max}$, and it is likely that both play a role in Bi2212 degradation.

Table 1: Performance of Bi2212 RWs Used in Quench Study [31]

| Sample | $I_c$ (A) | $J_c$ (A/mm$^2$) | $I_t$ (A) | $I_t/I_c$ | $J_{Ag}$ (A/mm$^2$) | MQE (J) | NZPV (cm/s) |
|---|---|---|---|---|---|---|---|
| Wire I, strand | 550 | 2334 | 450 | 81.8% | 1000 | 1.61 | 11.2 |
| Wire I, coil | 500 | 2122 | 400 | 80.0% | 909 | 2.38 | 3.6 |
| Wire II, strand | 500 | 4420 | 410 | 82.0% | 1284 | 0.78 | 17.4 |
| Wire II, coil | 360 | 3183 | 290 | 80.2% | 924 | 1.75 | 6.3 |

Table 2: Safe Operational Limits in Bi2212 RWs [31]

| Sample | $T_{max}$(K) | | $dT/dt|_{max}$(K/s) | | $dT/dx|_{max}$(K/cm) | |
|---|---|---|---|---|---|---|
| | Highest without degradation | Lowest with degradation | Highest without degradation | Lowest with degradation | Highest without degradation | Lowest with degradation |
| Wire I, strand | 317 | 350 | 695 | 700 | 95 | 100 |
| Wire I, coil | 330 | 358 | 770 | 802 | 82 | 93 |
| Wire II, strand | 193 | 200 | 592 | 605 | 65 | 66 |
| Wire II, coil | 139 | 167 | 219 | 255 | 43 | 48 |

*REBCO Coated Conductors*

The onset of degradation in a straight REBCO coated conductor was studied using a similar procedure as the Bi2212 study [20]. In this case, it was found that, when $T_{max}$ reached 490 K (±50 K), the conductor $I_c$ was degraded by ~5%. This corresponded to a strain of about 0.31%, which was ~60% of the irreversible strain reported for this conductor. In this experiment, $dT/dt|_{max}$ ~ 1800 K/s; values for $dT/dx|_{max}$ were as high as 27 K/mm, but the limited spatial resolution of the thermocouples implies that this value is probably lower than the actual temperature gradient in the conductor.

The microstructure of a degraded YBCO coated conductor was also studied via an etching technique that exposed the YBCO surface, facilitating microscopy of the surface [32]. In this study, the quenched sample experienced a 6.3% reduction in $I_c$. To expose the YBCO surface, the Cu stabilizer was first removed by a Cu etchant that contained S. Reactions between S and YBCO, seen after removal of the Ag cap layer, proved to be a signature of Ag/YBCO delamination (i.e., if the Ag/YBCO interface had not been breached during quenching, then the S would not have had access to the YBCO layer during etching of the Cu). This study showed clearly that pre-existing defects in the YBCO are the initiation points of subsequent degradation; two different defect structures were identified. At the edge of the conductor, where slitting occurs during conductor manufacturing, defects in the form of either the absence of Ag or delaminated Ag result in very high local electric field during a quench. This large, local electric field induces dendritic flux avalanches and a very high local temperature rise. The high temperature then causes further Ag delamination and a self-propagating effect which results in dendritic Ag delamination. The dendritic delamination in turn results in very narrow channels which facilitate Cu etchant penetration and a chemical reaction between S and YBCO. As a result, the microstructure shows a dendritic reaction pattern; EDS confirms that the dendrites have a high S content. This is seen in Fig. 3 and Fig. 4.

The other type of defect, labelled Q3 in Fig. 3, is observed within the YBCO layer but away from the conductor edge. This defect results in a reaction zone that shows significant S content. In this case, the Ag is both delaminated from the YBCO layer and breached such that Cu etchant penetrated and reacted with the underlying YBCO. As seen in Fig. 3 and Fig. 5, the degradation zone is nearly a perfect circle, indicative of current streamlines with a sharp boundary. An EDS map (Fig. 5) shows similar reactions as in the edge degradation, but also showed pure Ag particles (Fig. 6), which indicates that the local temperature was sufficient to create a Ag vapor that, upon cooling, results in such fine Ag precipitates.

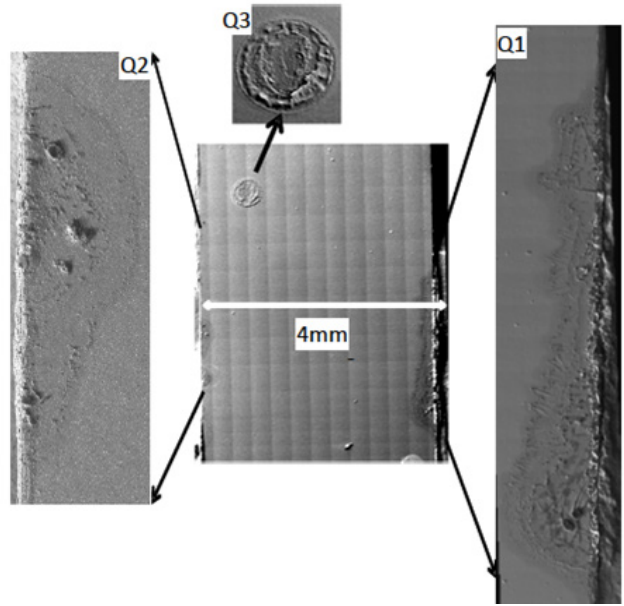

Figure 3: SEM micrographs of etched YBCO coated conductor, illustrating the two types of defects seen [32].

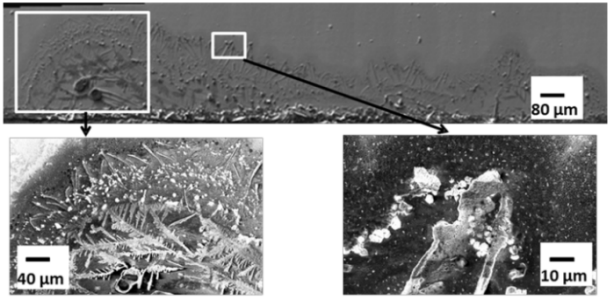

Figure 4: SEM micrographs of etched YBCO coated conductor, illustrating dendritic penetration from the edges [32].

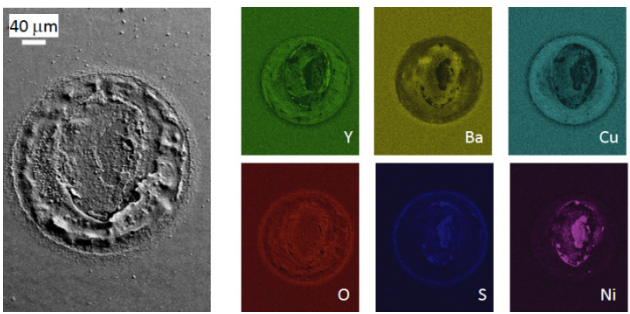

Figure 5: SEM micrographs and EDS maps of etched YBCO coated conductor, illustrating the degradation zone away from the conductor edge [32].

*Degradation Discussion*

Degradation in Bi2212 and REBCO conductors is quite different and generalized conclusions cannot be made. In fact, as both of these conductors are still under development, it is difficult to make quantitative conclusions that will be applicable to future conductors with improved properties. Nonetheless, some meaningful conclusions can be drawn. For Bi2212, the safe operational limits have large windows of uncertainty, which may be related to the inhomogeneous microstructure as well as differences in the way the wire was supported during the experiments. Bi2212 is mechanically relatively weak and thus improvements to operational limits may require significant improvements in the Bi2212 filament microstructure as well as the overall mechanical behavior of the conductor (e.g., a strengthened sheath [33]). For REBCO, there are also differences between different conductor types. Since the degradation is clearly initiated at manufacturing defects in the conductor, however, as the manufacturing processes improve it is likely that the conductor will become more quench resilient. One concern, however, is that that the susceptibility to degradation is related to the very high $J_c$ in the REBCO layer. Because these conductors have a low fill-factor, high $J_c$ in the REBCO layer is essential to their success.

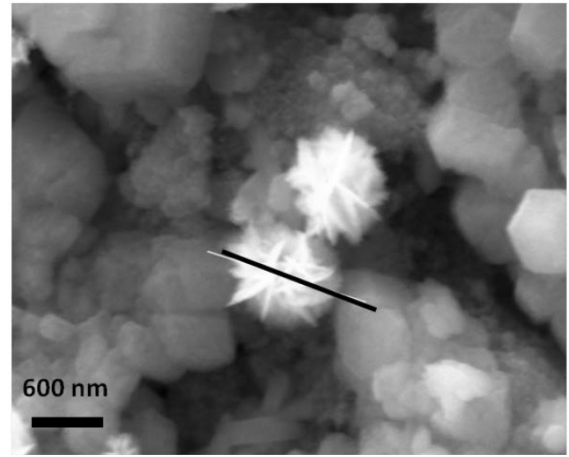

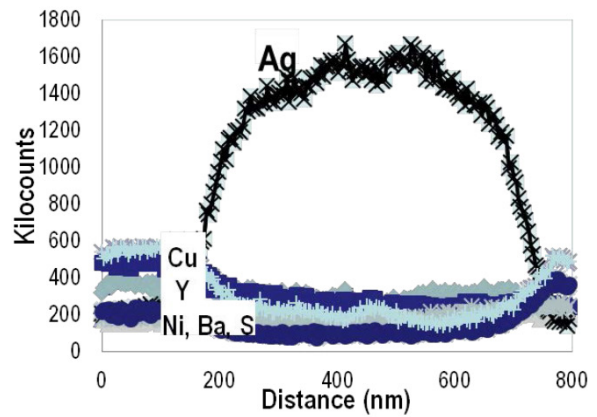

Figure 6: SEM micrograph within one of the rings in the reaction zone of Fig. 5, and an EDS line scan showing that this is a Ag precipitate [32].

## QUENCH DETECTION CHALLENGE

From the perspective of quench protection, the challenge of slow NZPV is quench detection; if the normal zone does not propagate quickly then neither does a detectable signal which may result in degradation before a protection system can take action. In principle, slow NZPV may also expand the time-budget, in which case the delayed detection would not be problematic as the two effects could cancel; i.e., relative to LTS magnets, everything would be similar but in "slow motion". The challenge arises, however, because traditional quench detection is based upon voltage measurements. Voltage, however, is simply the line integral of the electric field between the two voltage taps. The electric field profile is directly correlated with the temperature profile, so a normal zone with a short, highly peaked temperature profile produces the same voltage as a long normal zone with a relatively low peak temperature. The highly peaked temperature profile, however, which is what results from a slow NZPV, is more likely to induce degradation. Thus, either significantly enhanced quench propagation is required, or quench detection requires higher spatial resolution.

## Rayleigh scattering

One approach that has been proposed for distributed quench detection is Rayleigh scattering interrogation of an optical fiber that is co-wound with the conductor in the magnet [34]. While a variety of fiber-optic based sensing approaches have been studied for superconducting magnets [35-42], Rayleigh backscattering is the most versatile approach for sensing in HTS magnets because it relies on random fluctuations in the index profile along the optical fiber. Light sent down the fiber encounters these fluctuations, causing Rayleigh scattering. Thus, Rayleigh scattering uses naturally occurring defects present in even the best quality fibers as scattering points. For a given fiber, the scatter amplitude as a function of distance is a random but static property of that fiber and can be modeled as a long, weak Bragg grating with a random period. Thermally induced changes in the local period of the Rayleigh scatter pattern cause changes in the locally reflected spectrum and this spectral shift can be calibrated to form a distributed temperature sensor.

Recent work has confirmed that Rayleigh scattering, with the potential for spatial resolution well below 1 mm, allow fibers to be used as truly distributed sensors in HTS magnets [34]. The high spatial resolution, however, leads to a large amount of data to be analyzed in order to determine the spectral shift from a reference scan, which can lead to insufficient temporal resolution. Additionally, the length of fiber being monitored also increases the data volume, and thus the processing time can consume a large amount of the time quench protection time budget. Thus, an effective quench sensor based on Rayleigh scattering has an essential need for a balance between spatial resolution, temporal resolution, data analysis speed and monitoring length. To balance these needs, it is thus critical to understand the trade-off in spatial and temporal resolution that is required to protect HTS magnets.

## Modelling REBCO quench behaviour

To better understand all aspects of the dynamic quench behavior of REBCO conductors and magnets, an experimentally validated, multi-scale, two-dimensional/three-dimensional (2D/3D) model has been developed using the COMSOL platform. This model is able to accurately predict NZPV and MQE of conductors and magnets while also calculating the temperature, voltage and mechanical state of the various layers of materials that comprise the conductor [43-46]. Previously this model has been used to identify conductor variables that can influence quench behaviour [45]. Most recently, it has been used to identify the required spatial and temporal resolutions of a quench detection system for a YBCO coil that had been previously characterized experimentally [46]; this is illustrated in Fig. 7. This plot superimposes the requirements of the magnet for effective quench detection, based upon a minimum propagating zone criterion, with the capabilities of a distributed sensing system (in this example, a Rayleigh scattering system based upon optical backscatter reflectometry is used). Through this approach, the applicability of a sensor technology to a particular magnet can be assessed. Or, alternatively, the protectability of a magnet using a particular quench detection technology can be determined and the magnet redesigned as needed. This approach allows the quench detection system to be a consideration during the design of a magnet system rather than a response to the design.

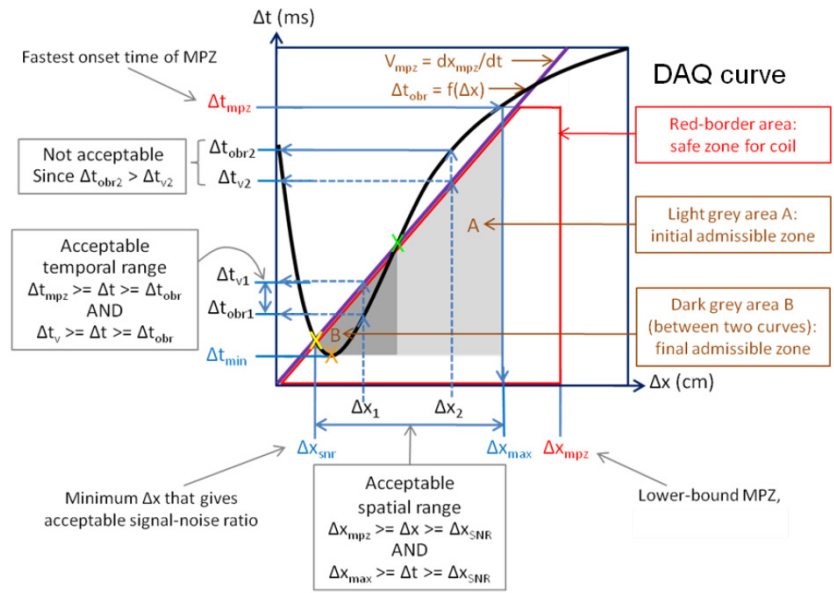

Figure 7: Graphical illustration of the required spatial and temporal resolutions for safe quench detection in a YBCO magnet. Superimposed upon the magnet requirements are the hypothetical capabilities of a Rayleigh-scattering based distributed sensor. The overlap then shows the safe operating range for the magnet with such a detection system. If there is no overlap, then the detection technology would not be applicable to the magnet [34].

# THERMALLY-CONDUCTING ELECTRICAL INSULATORS

Traditionally, quench propagation is much faster along the length of the conductor (longitudinal propagation) than from turn-to-turn (transverse propagation) because of the presence of turn-to-turn electrical insulation. An alternative approach to quench protection is to change this paradigm such that transverse propagation contributes significantly to the spreading of the normal zone, increasing the utilization of the available specific heat of the coil and reducing the rate at which the peak temperature increases. Although longitudinal normal zone propagation may be slower, the time budget for detection would be significantly increased.

The effect of thermally-conducting electrical insulator was demonstrated using the 2D/3D model of quench propagation. Fig. 8 shows the temperature versus time and location in YBCO coils insulated with (a) kapton and (b) a thermally conducting electrical insulator. The coil with the thermally conducting electrical insulator has a sixfold increase in MQE, a 50% reduction in peak temperature with a corresponding doubling of the end-to-end coil voltage. Thus, the magnet is more stable due to the accessibility of the specific heat of the entire coil, and a quench is more readily detected due to the higher voltage, while more resilient due to the reduced peak temperature.

While the results shown in Fig. 8 are the result of a computer model, this concept has been recently translated to practice for both Bi2212 and YBCO coils. Using a doped-titania insulation, the NZPV of both Bi2212 and YBCO coils has been increased by 275% [47]. This is illustrated in Fig. 9, which compares the voltage versus time of coils insulated with the doped titania insulation (left) and with traditional insulation (kapton for YBCO, braided mullite for Bi2212). Noting that the horizontal distance between the data represent the time delay for propagation, the effect of the insulation is clear. As this insulation system is improved further, magnets can evolve from linear, one-dimensional normal zones to (ultimately) spherical normal zones that maximize the coils ability to resist degradation.

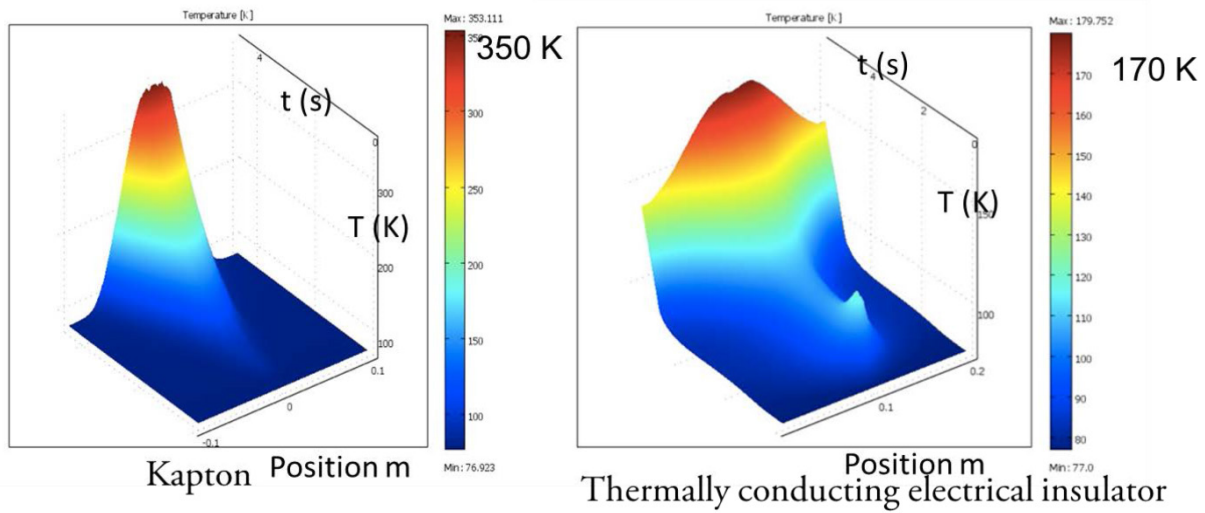

Figure 8: Comparison of the temperature profile of a YBCO magnet with (left) Kapton and (right) a thermally conducting electrical insulator [48].

# SUMMARY

The development of high field superconducting magnets using high temperature superconductors has progressed significantly due to advances in Bi2212 and YBCO conductor technologies. Quench propagation in HTS magnets, even at high magnetic fields, is very slow – as much as two orders of magnitude slower than LTS magnets – and thus quench protection remains a key hurdle before such magnets can be safely implemented.

In this paper a number of approaches to improving quench protection in HTS magnets have been summarized. One option is to improve the resilience of the conductors, thereby increasing the time-budget for detection and protection. This is likely to occur through the continuous evolution of Bi2212 conductors, but for REBCO coated conductors, which have very high $J_c$ in the superconducting layer, focusing specifically on the ability of a conductor to withstand a quench may be key.

Another approach is to implement a distributed quench detection sensor, such as an optical fiber monitored using Rayleigh scattering. Such a system has the potential for spatial resolution approaching the wavelength of the interrogating light, but at the expense of rapidly increasing data analysis times and thus a loss of temporal resolution. Understanding the required spatial and temporal resolutions for quench detection thus becomes essential.

Another option is transform quench propagation from being primarily a one-dimensional behavior to being a three-dimensional behavior. This may be achieved through the development of thermally-conducting electrical insulation that greatly reduces the peak temperature of the magnet relative to the coil voltage.

Quench behavior in HTS coils is a complex phenomenon that requires a dynamic understanding of both the conductor at the ~ μm scale and the magnet at the macroscopic scale. To facilitate this, a multi-scale model has been developed in COMSOL that allows complex analysis of REBCO magnets. This model can be effectively implemented to engineer the conductor, the magnet, and to understand the complex quench behavior.

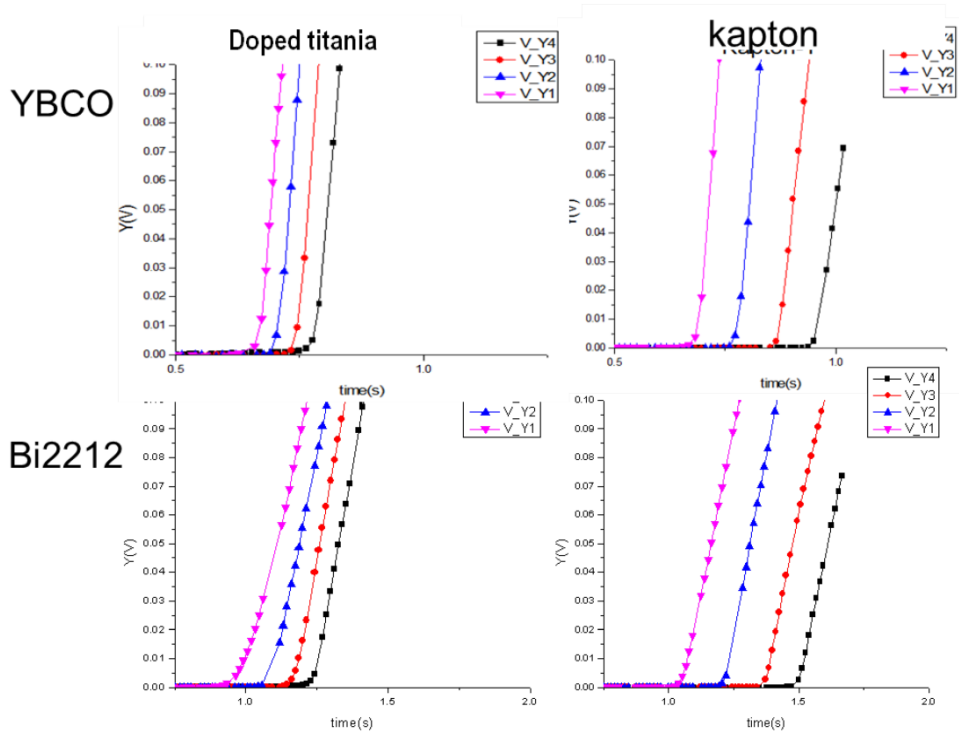

Figure 9: Illustration of the effect of the recently-developed thermally–conducing electrical insulator on quench propagation in YBCO and Bi2212 coils [47].

## ACKNOWLEDGMENT

The author is greatly indebted to and appreciative of a number of collaborators and students who have contributed to this work, including Wan Kan Chan, Davide Cruciani, Timothy Effio, Gene Flanagan, Andrew Hunt, Frank Hunte, Sasha Ishmael, Philippe Masson, Makita Phillips, Honghai Song, Ulf Trociewitz, Melanie Turenne, Xiaorong Wang, Marvis White and Liyang Ye.

## REFERENCES

[1] J. Schwartz, T. Effio, X. Liu, Q. V. Le, A. L. Mbaruku, H. J. Schneider-Muntau, T. Shen, H. Song, U. P. Trociewitz, X. R. Wang, and H. W. Weijers, High field superconducting solenoids via high temperature superconductors, IEEE Transactions on Applied Superconductivity, vol. 18, pp. 70-81, 2008.

[2] W. T. Nachtrab, C. V. Renaud, T. Wong, X. T. Liu, T. M. Shen, U. P. Trociewitz, and J. Schwartz, Development of high superconductor fraction $Bi_2Sr_2CaCu_2O_x$/Ag wire for MRI, IEEE Transactions on Applied Superconductivity, vol. 18, pp. 1184-1187, 2008.

[3] T. Shen, J. Jiang, A. Yamamoto, U. P. Trociewitz, J. Schwartz, E. E. Hellstrom, and D. C. Larbalestier, Development of high critical current density in multifilamentary round-wire Bi2Sr2CaCu2O8+delta by strong overdoping, Applied Physics Letters, vol. 95, Oct 2009.

[4] T. Shen, J. Jiang, F. Kametani, U. P. Trociewitz, D. C. Larbalestier, J. Schwartz, and E. E. Hellstrom, Filament to filament bridging and its influence on developing high critical current density in multifilamentary Bi2Sr2CaCu2Ox round wires, Superconductor Science & Technology, vol. 23, p. 025009, Feb 2010.

[5] K. R. Marken, H. P. Miao, M. Meinesz, B. Czabaj, and S. Hong, Progress in Bi-2212 wires for high magnetic field applications, IEEE Transactions on

Applied Superconductivity, vol. 16, pp. 992-995, Jun 2006.
[6] H. Miao, K. R. Marken, M. Meinesz, B. Czabaj, and S. Hong, Development of round multifilament Bi-2212/Ag wires for high field magnet applications, IEEE Transactions on Applied Superconductivity, vol. 15, pp. 2554-2557, Jun 2005.
[7] W. T. Nachtrab, X. T. Liu, T. Wong, and J. Schwartz, Effect of Solidification Conditions on Partial Melt Processed Bi2212/Ag Round Wire, IEEE Transactions on Applied Superconductivity, vol. 21, pp. 2795-2799, Jun 2011.
[8] H. W. Weijers, U. P. Trociewitz, K. Marken, M. Meinesz, H. Miao, and J. Schwartz, The generation of 25.05 T using a 5.11 T $Bi_2Sr_2CaCu_2O_x$ superconducting insert magnet, Superconductor Science & Technology, vol. 17, pp. 636-644, Apr 2004.
[9] W. D. Markiewicz, H. W. Weijers, P. D. Noyes, U. P. Trociewitz, K. W. Pickard, W. R. Sheppard, J. J. Jaroszynski, A. Xu, D. C. Larbalestier, and D. W. Hazelton, 33.8 Tesla with a $YBa_2Cu_3O_{7-x}$ superconducting test coil, in Advances in Cryogenic Engineering, Vols 55a and 55b. vol. 1218, J. G. Weisend, J. Barclay, S. Breon, J. Demko, M. DiPirro, J. P. Kelley, P. Kittel, A. Klebaner, J. Marquardt, G. Nellis, T. Peterson, J. Pfotenhauer, S. VanSciver, M. Zagarola, and A. Zeller, Eds., ed, 2010, pp. 225-230.
[10] R. A. Badcock, N. J. Long, M. Mulholland, S. Hellmann, A. Wright, and K. A. Hamilton, Progress in the Manufacture of Long Length YBCO Roebel Cables, IEEE Transactions on Applied Superconductivity, vol. 19, pp. 3244-3247, Jun 2009.
[11] D. C. van der Laan, X. F. Lu, and L. F. Goodrich, Compact $GdBa_2Cu_3O_{7-d}$ coated conductor cables for electric power transmission and magnet applications, Superconductor Science & Technology, vol. 24, Apr 2011.
[12] F. Trillaud, A. Caruso, J. Barrow, B. Trociewitz, U. P. Trociewitz, H. W. Weijers, and J. Schwartz, Normal Zone Generation and Propagation in YBa2Cu3O7-? Coated Conductors Initialized by Localized Pulsed Disturbances, Advances in Cryogenic Engineering Materials, 2004, pp. 852-859.
[13] F. Trillaud, H. Palanki, U. P. Trociewitz, S. H. Thompson, H. W. Weijers, and J. Schwartz, Normal zone propagation experiments on HTS composite conductors, Cryogenics, vol. 43, pp. 271-279, Mar-May 2003.
[14] H. H. Song and J. Schwartz, Stability and Quench Behavior of YBa2Cu3O7-x Coated Conductor at 4.2 K, Self-Field, IEEE Transactions on Applied Superconductivity, vol. 19, pp. 3735-3743, Oct 2009.
[15] C. L. H. Thieme, K. Gagnon, H. Song, and J. Schwartz, Stability of second generation HTS pancake coils at 4.2 K for high heat flux applications, IEEE Transactions on Applied Superconductivity, vol. 19, p. 1626-1632 2009.
[16] H. Song, K. Gagnon, and J. Schwartz, Quench behavior of conduction-cooled $YBa_2Cu_3O_{7-\delta}$ coated-conductor pancake coils stabilized with brass and copper, Superconductor Science & Technology, vol. 23, p. 10, 2010.
[17] X. R. Wang, A. R. Caruso, M. Breschi, G. M. Zhang, U. P. Trociewitz, H. W. Weijers, and J. Schwartz, Normal zone initiation and propagation in Y-Ba-Cu-O coated conductors with Cu stabilizer, IEEE Transactions on Applied Superconductivity, vol. 15, pp. 2586-2589, Jun 2005.
[18] X. Wang, U. P. Trociewitz, and J. Schwartz, Near-adiabatic quench experiments on short $YBa_2Cu_3O_{7-d}$ coated conductors, Journal of Applied Physics, vol. 101, p. 053904, Mar 2007.
[19] X. R. Wang, U. P. Trociewitz, and J. Schwartz, Self-field quench behavior of YBCO coated conductors with different stabilizers, Superconductor Science & Technology, vol. 22, 2009.
[20] X. Wang, U. P. Trociewitz, and J. Schwartz, Critical current degradation of short YBa2Cu3O7-s coated conductor due to an unprotected quench, Superconductor Science & Technology, vol. 24, p. 035006, 2011.
[21] W. S. Kim, F. Trillaud, M. C. Ahn, Y. Iwasa, X. Peng, and M. Tomsic, Normal zone propagation in 2-dimensional YBCO winding pack models, IEEE Transactions on Applied Superconductivity, vol. 18, pp. 1249-1252, June 2008.
[22] L. A. Angurel, E. Martinez, J. Pelegrin, R. Lahoz, G. F. de la Fuente, N. Andres, M. P. Arroyo, Y. Y. Xie, and V. Selvamanickam, Changes in the Thermal Stability of 2G HTS Wires by Local Modification of the Stabilization Layer, IEEE Transactions on Applied Superconductivity, vol. 21, pp. 3017-3020, Jun 2011.
[23] F. Trillaud, M. C. Ahn, J. Bascunan, W. S. Kim, J. P. Voccio, and Y. Iwasa, Quench behavior, quench protection of a YBCO test coil assembly, IEEE Transactions on Applied Superconductivity, vol. 18, pp. 1329-1332, June 2008.
[24] T. Effio, U. P. Trociewitz, X. Wang, and J. Schwartz, Quench induced degradation in $Bi_2Sr_2CaCu_2O_{8+x}$ tape conductors at 4.2 K, Superconductor Science & Technology, vol. 21, Apr 2008.
[25] U. P. Trociewitz, B. Czabaj, S. Hong, Y. Huang, D. C. Knoll, D. C. Larbalestier, W. D. Markiewicz, H. Miao, M. Meinesz, X. Wang, and J. Schwartz, Quench studies on a layer-wound $Bi_2Sr_2CaCu_2O_x$/AgX coil at 4.2 K, Superconductor Science & Technology, vol. 21, Feb 2008.
[26] L. Ye, F. Hunte, and J. Schwartz, Effects of high magnetic field on the low-temperature quench behavior of $Bi_2Sr_2CaCu_2O_x$ coils, Superconductor Science & Technology, vol. 26, p. 055006, 2013.
[27] C. V. Varanasi, J. Burke, H. Wang, J. H. Lee, and P. N. Barnes, Thick $YBa_2Cu_3O_{7-x}$+$BaSnO_3$ films with enhanced critical current density at high magnetic

fields, Applied Physics Letters, vol. 93, p. 3, Sep 2008.
[28] V. Selvamanickam, Y. Chen, X. Xiong, Y. Xie, X. Zhang, A. Rar, M. Martchevskii, R. Schmidt, K. Lenseth, and J. Herrin, Progress in second-generation HTS wire development and manufacturing, Physica C, vol. 468, pp. 1504-1509, 2008.
[29] T. Aytug, M. Paranthaman, K. J. Leonard, K. Kim, A. O. Ijaduola, Y. Zhang, E. Tuncer, J. R. Thompson, and D. K. Christen, Enhanced flux pinning and critical currents in YBa2Cu3O7-delta films by nanoparticle surface decoration: Extension to coated conductor templates, Journal of Applied Physics, vol. 104, p. 6, Aug 2008.
[30] S. H. Wee, A. Goyal, Y. Li, Y. L. Zuev, S. Cook, and L. Heatherly, The incorporation of nanoscale columnar defects comprised of self-assembled $BaZrO_3$ nanodots to improve the flux pinning and critical current density of $NdBa_2Cu_3O_{7-d}$ films grown on RABiTS, Superconductor Science & Technology, vol. 20, pp. 789-793, Aug 2007.
[31] L. Ye, D. Cruciani, T. Effio, F. Hunte, and J. Schwartz, On the causes of degradation in $Bi_2Sr_2CaCu_2O_{8+x}$ round wires and coils by quenching at 4.2 K, IEEE Transactions on Applied Superconductivity, submitted 2013.
[32] H. H. Song, F. Hunte, and J. Schwartz, On the role of pre-existing defects and magnetic flux avalanches in the degradation of $YBa_2Cu_3O_{7-x}$ coated conductors by quenching, Acta Materialia, vol. http://dx.doi.org/10.1016/j.actamat.2012.09.003, 2012.
[33] A. Kajbafvala, W. Nachtrab, X. F. Lu, F. Hunte, X. Liu, N. Cheggour, T. Wong, and J. Schwartz, Dispersion-Strengthened Silver Alumina for Sheathing $Bi_2Sr_2CaCu_2O_{8+x}$ Multifilamentary Wire, IEEE Transactions on Applied Superconductivity, vol. 22, pp. 8400210-8400210, 2012.
[34] W. K. Chan, G. Flanagan, and J. Schwartz, Spatial and temporal resolution requirements for quench detection in (RE)Ba2Cu3Ox magnets using Rayleigh-scattering based fiber optics distributed sensing, Superconductor Science & Technology, submitted 2013
[35] F. Hunte, H. Song, J. Schwartz, R. P. Johnson, and M. Turenne, Fiber Bragg Optical Sensors for YBCO Applications, presented at the 2009 Particle Accelerator Conference, Vancouver, Canada, 2009.
[36] M. Turenne, R. P. Johnson, F. Hunte, and J. Schwartz, Multi-Purpose Fiber Optic Sensors for High Temperature Superconductor Magnets, presented at the 2009 Particle Accelerator Conference, Vancouver, Canada, 2009.
[37] S. B. Mahar, Spontaneous Brillouin Scattering Quench Diagnostics for Large Superconducting Magnets, PhD., Department of Nuclear Science & Engineering, Massachusetts Institute of Technology, Cambridge, 2008.
[38] S. Pourrahimi, S. P. Smith, J. H. Schultz, and J. V. Minervini, Performance of the US quench detection systems in the QUELL experiments, IEEE Transactions on Applied Superconductivity, vol. 7, pp. 447-450, 1997.
[39] R. RajiniKumar, K. G. Narayankhedkar, G. Krieg, M. Susser, A. Nyilas, and K. P. Weiss, Fiber Bragg gratings for sensing temperature and stress in superconducting coils, 2006, pp. 1737-1740.
[40] J. H. Schultz, S. Pourrahimi, S. Smith, and P. W. Wang, Principles of advanced quench detection design in cable-in-conduit (CICC) magnets, IEEE Transactions on Applied Superconductivity, vol. 7, pp. 455-460, 1997.
[41] J. M. van Oort and R. M. Scanlan, A fiber-optic strain measurement and quench localization system for use in superconducting accelerator dipole magnets, IEEE Transactions on Applied Superconductivity, vol. 5, pp. 882-885, 1995.
[42] J. M. van Oort and H. H. J. ten Kate, A fiber-optic sensor for strain and stress measurement in superconducting accelerator magnets, IEEE Transactions on Magnetics, vol. 30, pp. 2600-2603, 1994.
[43] W. K. Chan, P. J. Masson, C. Luongo, and J. Schwartz, "Three-dimensional micrometer-scale modeling of quenching in high aspect ratio $YBa_2Cu_3O_{7-\delta}$ coated conductor tapes. Part I: Model development and validation," *IEEE Transactions on Applied Superconductivity,* vol. **20**, pp. 2370-2380, 2010.
[44] W. K. Chan, P. J. Masson, C. A. Luongo, and J. Schwartz, Influence of inter-layer contact resistances on quench propagation in $YBa_2Cu_3O_x$ coated conductors, IEEE Transactions on Applied Superconductivity, vol. 19, p. 2490—2495, 2009.
[45] W. K. Chan and J. Schwartz, Three-Dimensional Micrometer-Scale Modeling of Quenching in High-Aspect-Ratio $YBa_2Cu_3O_{7-d}$ Coated Conductor Tapes-Part II: Influence of Geometric and Material Properties and Implications for Conductor Engineering and Magnet Design, IEEE Transactions on Applied Superconductivity, vol. 21, pp. 3628-3634, Dec 2011.
[46] W. K. Chan and J. Schwartz, A hierarchical, three-dimensional, multiscale electro-magneto-thermal model of quenching in $REBa_2Cu_3O_{7-d}$ coated conductor based coils, IEEE Transactions on Applied Superconductivity, vol. 22, p. 4706010, 2012.
[47] S. Ishmael, H. Lee, M. White, F. Hunte, X. T. Liu, N. Mandazy, J. Muth, G. Naderi, L. Ye, A. Hunt, and J. Schwartz, Enhanced quench propagation in $Bi_2Sr_2CaCu_2O_x$ and $YBa_2Cu_3O_{7-d}$ coils via nanoscale, doped-titania based, thermally-conducting electrical insulator, IEEE Transactions on Applied Superconductivity, 2013 (in press).
[48] P. Masson, 2007.